\documentclass[prl,twocolumn,aps,showpacs]{revtex4}
\usepackage{graphicx}

\begin{document}

\title{Electron-hole Asymmetry and Quantum Critical Point in
Hole-doped BaFe$_2$As$_2$}

\author{Gang Xu, Haijun Zhang, Xi Dai and Zhong Fang}

\affiliation{Beijing National Laboratory for Condensed Matter Physics,
and Institute of Physics, Chinese Academy of Sciences, Beijing 100190,
China;}

\date{\today}

\begin{abstract}
We show, from first-principles calculations, that the hole-doped side
of FeAs-based compounds is different from its electron-doped
counterparts. The electron side is characterized as Fermi surface
nesting, and SDW-to-NM quantum critical point (QCP) is realized by
doping. For the hole-doped side, on the other hand, orbital-selective
partial orbital ordering develops together with checkboard
antiferromagnetic (AF) ordering without lattice distortion. A unique
SDW-to-AF QCP is achieved, and $J_2$=$J_1/2$ criteria (in the
approximate $J_1\&J_2$ model) is satisfied. The observed
superconductivity is located in the vicinity of QCP for both sides.
\end{abstract}

\maketitle

The superconductivity found in FeAs-layer-based compounds\cite{LOFS}
is attractive and challenging not only because it is the only
non-cuprate system, which shows superconductivity beyond 40
K~\cite{SOFS}, but also because of the magnetic nature of Fe, and of
the multi-band character of the system. Initial studies concentrated
on the electron-doped compounds, such as
$Re$(O$_{1-x}$F$_x$)FeAs~\cite{LOFS} or
$Re$O$_{1-\delta}$FeAs~\cite{Oxygen}.  For the parent compound
LaOFeAs, it was first pointed out by Singh et.al~\cite{Singh} and
G. Xu et. al.~\cite{fang}, based on first-principles calculations,
that magnetic instabilities play crucial roles for the understanding
of superconductivity, and LaOFeAs is located at the border line of
both antiferromagnetic (AF) and ferromagnetic (FM) instabilities. Soon
later it was realized that significant Fermi surface nesting exists
between the hole and electron Fermi surfaces (FS) connected by a
$q$=($\pi,\pi$) vector~\cite{Mazin,SDW}. Based on transport and
optical measurements and detailed first-principles calculations,
J. Dong et.al~\cite{SDW} proposed that a stripe-type spin-density-wave
(SDW) state should be stabilized at low temperature, and the essential
physics to be discussed here is the competing orders between SDW and
superconducting states. The predicted SDW state was confirmed by
neutron experiment, and the stripe-type spin ordering pattern was
observed~\cite{Neutron}.  The success of LDA or GGA-type
calculations~\cite{Yil,Tera} for the present systems are in sharp
contrast to the failure when applied to cuprates, although the
electron-correlation may play certain roles~\cite{kotliar1}.  The
metallic nature of the SDW state also suggests that the FS nesting is
an important issue to understand the physics. The observed small
ordered moment~\cite{Neutron} (compared with first-principles
calculations) remains to be a difficult issue, however it was
demonstrated by recent careful study~\cite{Yil,Tera,Mazin2} that the
discrepancy can be much reduced if the optimized structure is used.

The recent interests however move to the hole-doped
Ba$_{1-x}$K$_x$Fe$_2$As$_2$ or Sr$_{1-x}$K$_x$Fe$_2$As$_2$ with Tc up
to 38K~\cite{BFS-SC}, because very large single crystal can be
synthesized~\cite{BFS-single}. For the parent compound BaFe$_2$As$_2$,
the SDW instability similar to LaOFeAs was observed~\cite{BFS-SDW},
and it was demonstrated by first-principles
calculations~\cite{BFS-cal} that BaFe$_2$As$_2$ is electronically
similar to LaOFeAs. However, as we will show in this paper, the
situation is true for the parent compounds, but the physics at the
hole-doping side is quite different with the undoped or electron-doped
side.

We performed systematic first principles calculations for both
hole-doped Ba$_{1-x}$K$_x$Fe$_2$As$_2$ and electron-doped
LaO$_{1-x}$F$_x$FeAs, using the virtual crystal approximation for the
doping. We used the experimental lattice parameters with the optimized
internal coordinates of As, as suggested by previous
studies~\cite{Mazin2}. Two different antiferromagnetic ordered
structures (as defined in Fig.1) are considered in the present study.

\begin{figure}
\includegraphics[clip,scale=0.45]{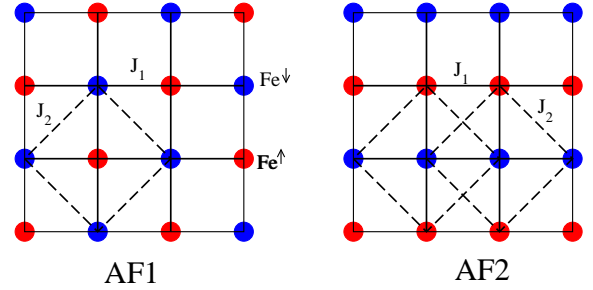}
\caption{The magnetic ordering patterns of two kinds of
antiferromagnetic states. The checkboard ordering is called AF1, and
the strip-type SDW state~\cite{SDW} is called as AF2 here. Two
exchange coupling constants, $J_1$ and $J_2$, are defined for the
nearest neighbor and next-nearest neighbor interactions,
respectively.}
\end{figure}

\begin{figure}
\includegraphics[clip,scale=0.12]{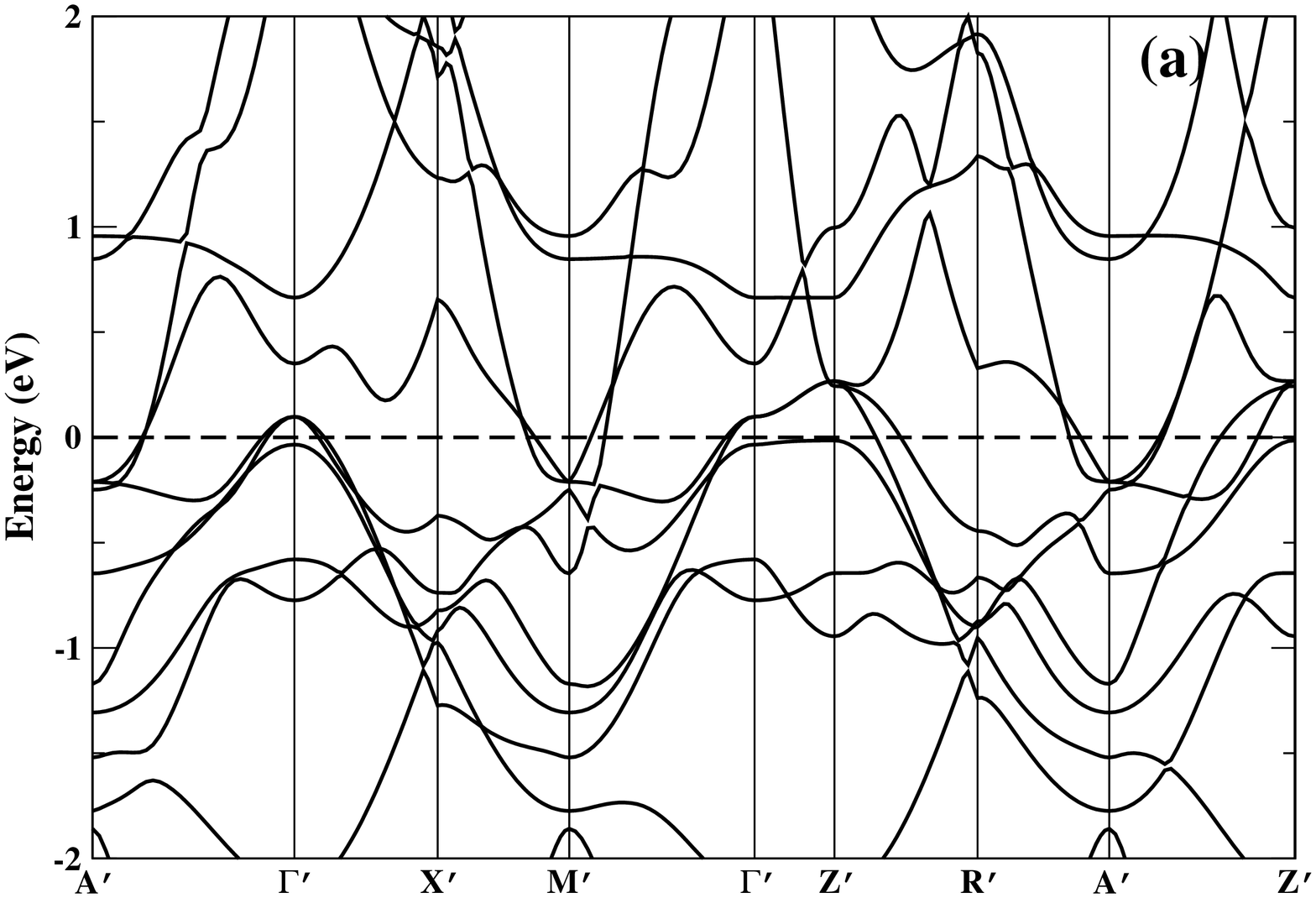}
\includegraphics[clip,scale=0.24]{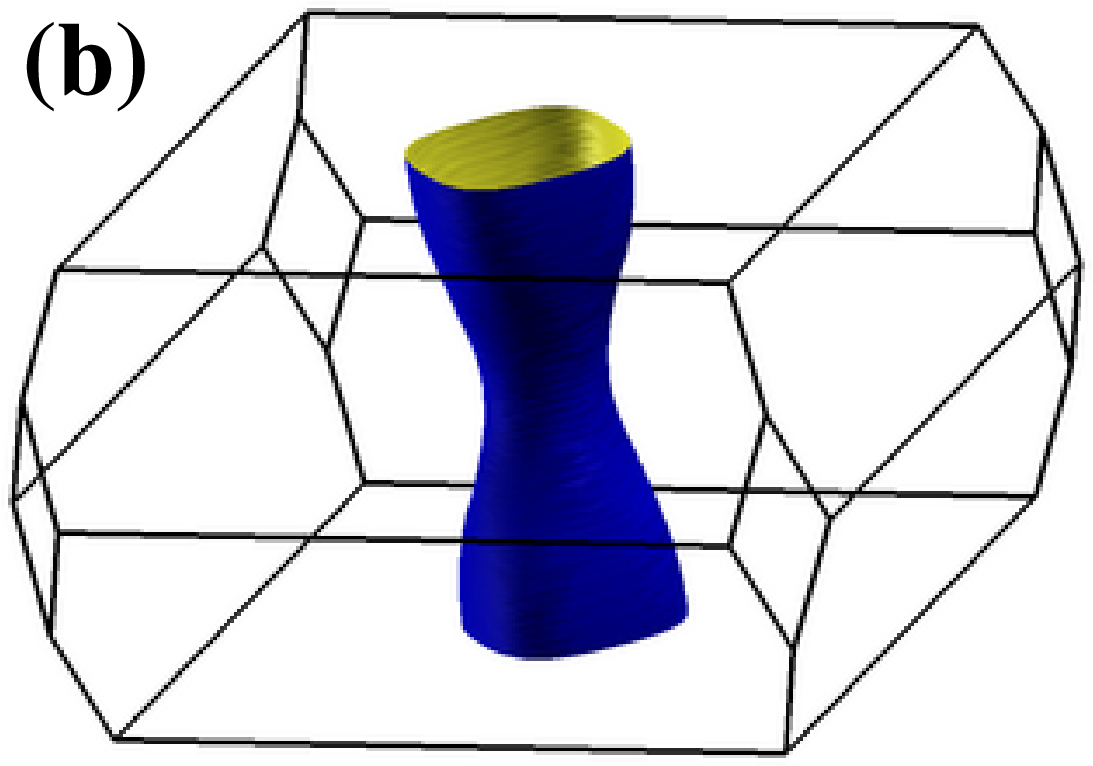}
\includegraphics[clip,scale=0.24]{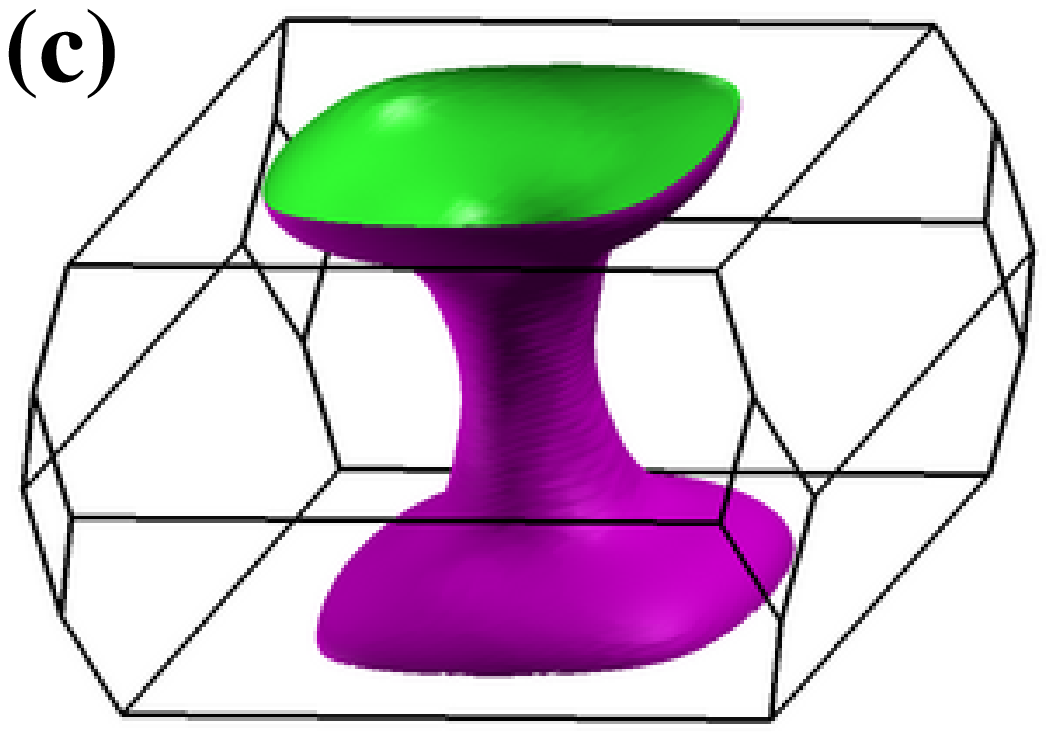} \\
\includegraphics[clip,scale=0.24]{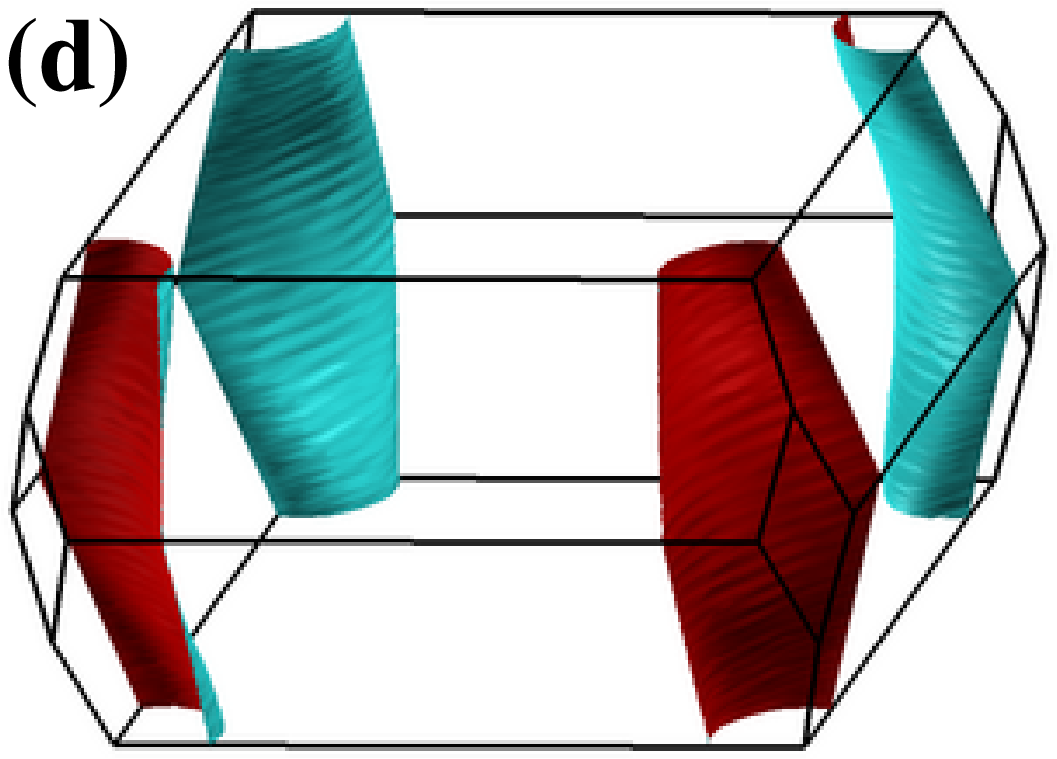}
\includegraphics[clip,scale=0.24]{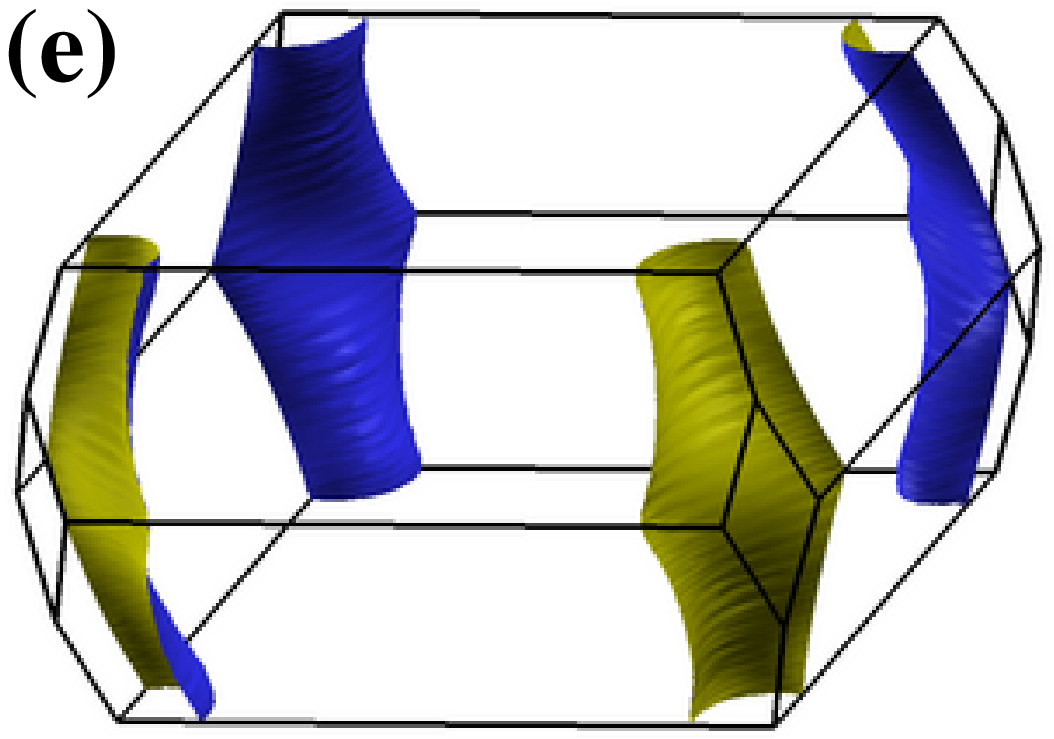}
\includegraphics[clip,scale=0.24]{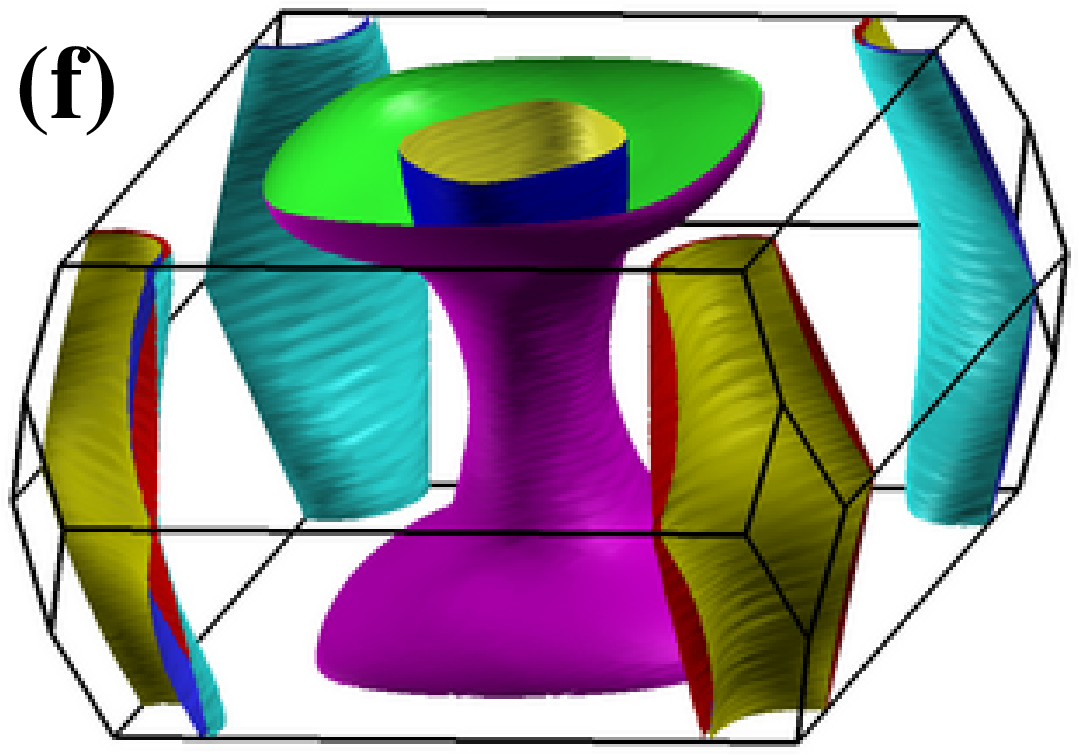}
\includegraphics[clip,scale=0.5]{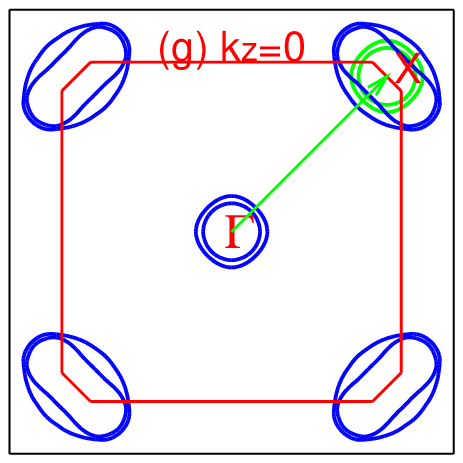} \ \ \
\includegraphics[clip,scale=0.5]{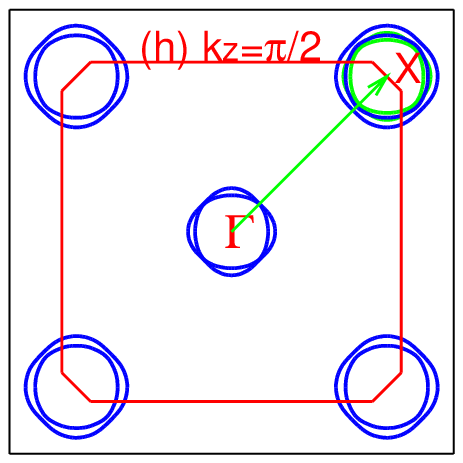} \ \ \
\includegraphics[clip,scale=0.5]{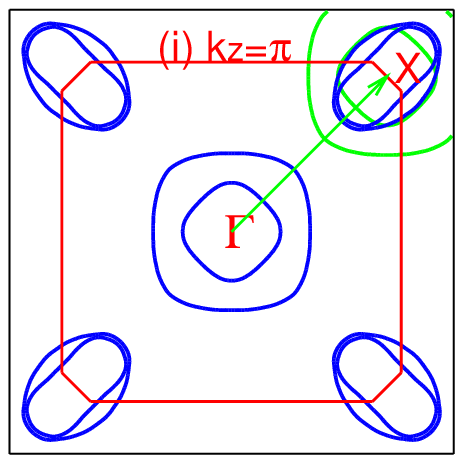}
\includegraphics[clip,scale=0.46]{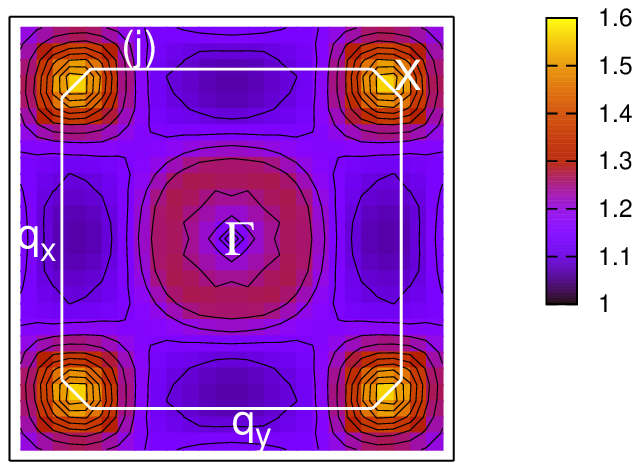}\ \  \  \
\includegraphics[clip,scale=0.13]{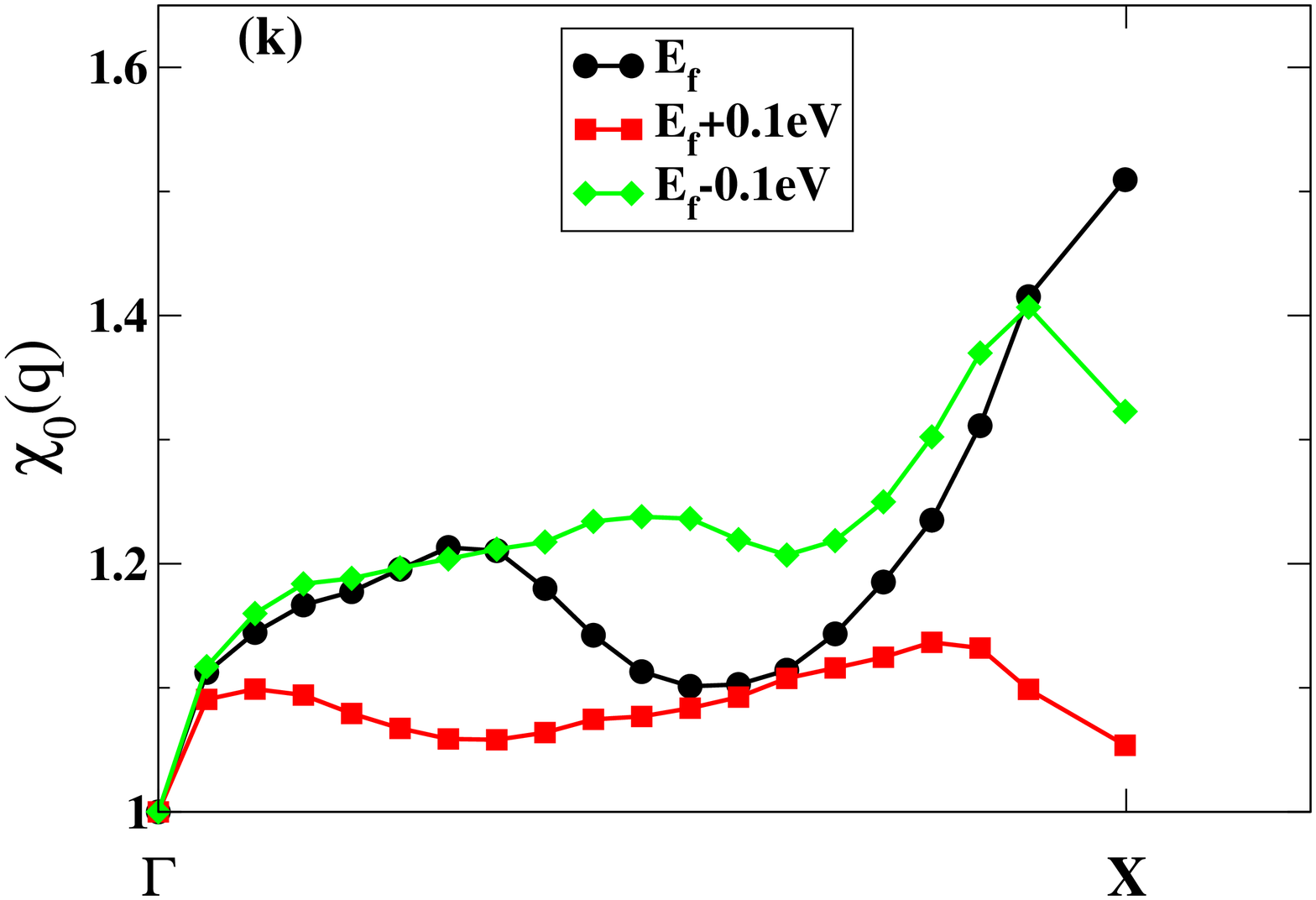}
\caption{The calculated electronic properties of BaFe$_2$As$_2$: (a)
The band structure plotted in the Brillouin Zone (BZ) of LaOFeAs
structure; (b)-(f) The Fermi surfaces (FS), and (f) is the plot for
total five FS; (g)-(i) The cuted FS for fixed $k_z$=0, $\pi/2$ and
$\pi$ planes. (j)-(k) The calculated Lindhard response function
$\chi_0(q)$ for the $q_z$=0 plane and for the $\Gamma$-X line. The X
point of BaFe$_2$As$_2$ BZ corresponds to the M point of LaOFeAs BZ.}
\end{figure}

{\it Parent compound BaFe$_2$As$_2$}: Fig.2 shows the calculated
electronic properties of BaFe$_2$As$_2$ in the non-magnetic (NM)
solution. Similar to LaOFeAs, there are two circle-like hole-type FS
around the $\Gamma$ point, and two ellipse-like electron-type FS around
the X point. The X point here corresponds to the M point in the
Brillouin Zone (BZ) of LaOFeAs, because BaFe$_2$As$_2$ has
body-centered tetragonal structure. On the other hand, different with
LaOFeAs, the band dispersion along the $z$ direction is stronger. In
particular, as shown in Fig.2 (g)-(i), for the $k_z$=0 ($k_z$=$\pi$)
plane, the size of hole-pockets is smaller (larger) than the
electron-pockets, and only for the $k_z$=$\pi$/2 plane, the size of
two are almost equal. This suggests the enhanced three-dimensionality
in BaFe$_2$As$_2$. The fact that the electron pockets have different
orientations for different $k_z$ plane can help us to understand the
ARPES results, and indeed the calculated FS shape can be well compared
with the recent ARPES measurement~\cite{ARPES}, again suggesting the
quality of first-principles calculations.

In addition to the FS characters, we calculated the Lindhard response
function $\chi_0(q)$ as shown in Fig.2 (j)-(k). Also similar to
LaOFeAs, we found significant FS nesting for the $q$=($\pi$, $\pi$)
vector, and the nesting can be suppressed by either electron or hole
doping. As the results, the stripe-type SDW state (called AF2 as shown
in Fig.1) similar to LaOFeAs is stabilized for the ground state (see
Fig.3 for the stabilization energy). The calculated moment is about
1.5 $\mu_B$/Fe, again larger than the measure moment of
0.8$\mu_B$/Fe~\cite{BFS-SDW}. All those facts suggest that the
Ba$Fe_2$As$_2$ is very similar to LaOFeAs except the enhanced
three-dimensionality.

{\it Effect of doping}: The electron and hole doped compounds are
however quite different as will be addressed in this part. Fig.3 shows
the calculated total energies of AF1 and AF2 with respect to the NM
solution, and the corresponding magnetic moment for both
Ba$_{1-x}$K$_x$Fe$_2$As$_s$ and LaO$_{1-x}$F$_x$FeAs. The qualitative
behavior as function of doping is definitely very different. We start
the discussion from LaOFeAs side. The ground state of LaOFeAs is AF2
as suggested before~\cite{SDW} and confirmed by
neutron~\cite{Neutron}, in the presence of Fermi surface nesting. By
the electron-doping, the nesting effect is suppressed and the
stabilization energy of AF2 is reduced. It is also shown in Fig.3 that
the size of the ordered moment is reduced by electron-doping. If the
doping $x$ is approaching 0.2, the AF2 state and its moment will be
totally suppressed, and the AF2-to-NM quantum critical point (QCP) is
realized~\cite{QCP}.

\begin{figure}
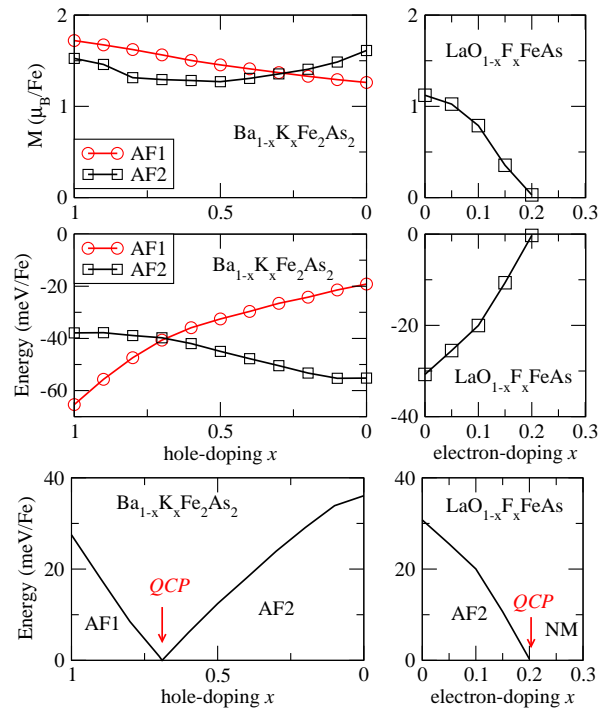

\includegraphics[clip,scale=0.45]{E_x.eps}
\includegraphics[clip,scale=0.45]{phase-3.eps}
\caption{The calculated magnetic moment and stabilization energies of
AF1 and AF2 states in Ba$_{1-x}$K$_x$Fe$_2$As$_2$ and
LaO$_{1-x}$F$_x$FeAs, with respect to the NM solution (energy zero
line). The lower panels show the ground state phase diagram
constructed from total energy calculations. Note the two different
quantum critical points (QCP).}
\end{figure}

Although the BaFe$_2$As$_2$ itself is quite similar to LaOFeAs,
however, with the hole-doping, the system behaviors quite
differently. First, the energy gains of both AF1 and AF2 relative to
NM state are substantial and it is not suppressed significantly by
hole-doping. Second, the calculated moment is almost constant and do
not reduce with doping. What is even more interesting is that the AF1
and AF2 states become degenerate at about $x$=0.7, and the AF1 state
is more stable than AF2 beyond this doping. The AF1-to-AF2 QCP is
therefore realized by hole-doping in sharp contrast to the
electron-doped side.  The constructed ground state phase diagram is
shown in Fig.3. The electron and the hole-doped sides are asymmetric,
and they are physically quite different: (1) the AF2-to-NM QCP is
realized for the electron-doped side, while AF2-to-AF1 QCP is realized
for the hole-doped side; (2) the electron-doped side can be understood
from FS nesting, however the hole-doped side behaviors more or less
like localized electron.

\begin{figure}[tbp]
\includegraphics[clip,scale=0.45]{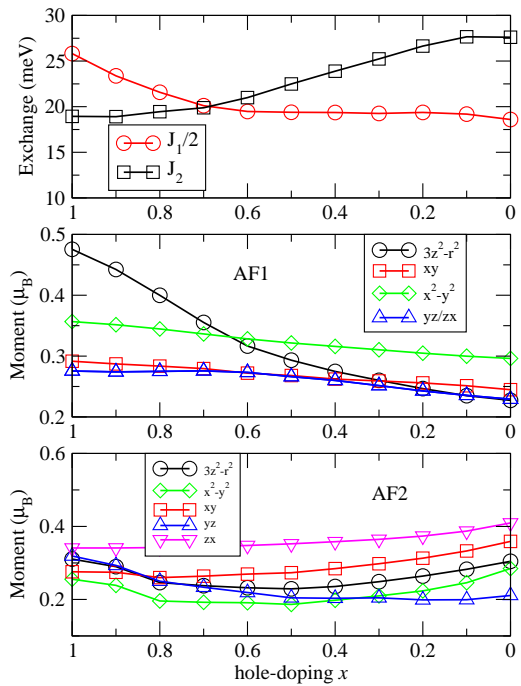}
\caption{Upper panel: The estimated exchange couplings ($J_1$ and
$J_2$) as function of hole-doping $x$ in
Ba$_{1-x}$K$_x$Fe$_2$As$_2$. Middle and lower panels: The calculated
magnetic moment for each 3$d$ orbitals for AF1 and AF2 states
respectively.}
\end{figure}

The first question to be answered at this stage is why the hole-doped
side behaviors so differently with the electron-doped side?  A simple
understanding is to look at the density of state (DOS) of parent
compounds, which is asymmetric around Fermi level $E_f$. It is
characterized as strong peak and high DOS below $E_f$, however
dispersive bands (mostly $xy$) and low DOS above $E_f$. By
electron-doping the electron pockets formed by the wide $xy$ band
dominates, and the system becomes more and more itinerant. On the
other hand, by hole-doping, the $E_f$ is shifted into the narrow band
region with high DOS peak. The effective $U_{eff}$(=$U/W$) for those
flat bands are therefore larger than the electron-doped side. In
addition, we should also notice that the number of $d$ electrons,
which is 6/Fe for undoped compounds, is reduced to be more close to
5/Fe (half-filled) by hole-doping.

The second question is what is the difference between the AF1 and the
AF2 states? The simple answer is that orbital-selective partial
orbital ordering develops in the AF1 state. For this purpose, we plot
in Fig.4 the calculated magnetic moment decoupled to each orbitals of
Fe. For the AF2 solution, the magnetic moment is nearly uniformly
distributed to each orbitals. (The difference between $yz$ and $zx$
orbitals is due to the stripe-ordering, which breaks the two-fold
rotational symmetry). However, the AF1 solution is unique: one of the
orbitals, 3$z^2-r^2$, which forms a narrow band below $E_f$,
contributes to the magnetic moment dominantly. Clearly the
multi-orbital physics and the orbital degrees of freedom play crucial
role here. Finally, at the x=1.0, the partial orbital ordering of
$3z^2-r^2$ develops in the AF1 state. With the hole-doping, it is
therefore realized that the orbital-selective magnetic transition
involves (the magnetization is transfered from other orbitals to the
$3z^2-r^2$ orbital).

The third question is why the AF1 is stabilized for large hole-doping
side (relative to AF2 state)?  This can be understood from the
$J_1$-$J_2$ model proposed previously for LaOFeAs~\cite{Yil}. For the
strong coupling fully-localized limit (localized spin system), the
$J_1$-$J_2$ model can be used to describe the possible ordered states
as shown in Fig.1. Here $J_1$ is the nearest-neighbor exchange
coupling, and $J_2$ is the next-nearest-neighbor exchange coupling. The
positive sign is defined for AF coupling. From the mean field solution
of this model, the total energy of AF1 state is given as
$E_1$=-2$J_1$+2$J_2$, and the energy of AF2 state is given as
$E_2$=-2$J_2$ per Fe. Therefore if $J_2$=$J_1$/2, the two ordered
state will be energetically degenerate, and strong quantum fluctuation
will be expected near this critical point. Such model has been suggested
to describe the physical properties of LaOFeAs, however,
first-principles calculations~\cite{Yil,Tera} suggest that the
electron-doped LaOFeAs system is actually far away from the
degeneracy, namely $J_2>>J_1$/2. On the other hand, here we show that
the quantum degeneracy can be actually realized in the hole-doped
side. Of course, our system is still away from the fully localized
insulating region (therefore the usage of $J_1$-$J_2$ model here is
not rigorously justified). However, We can borrow the simple idea of
$J_1$-$J_2$ model and qualitative understand the physics.  We map the
calculated total energy to this model, and estimate the $J_1$ and
$J_2$ coupling constants as shown in Fig.4. (The spin $S$=1 is used as
suggested from the calculated magnetic moment). Indeed, the
$J_2$=$J_1$/2 criteria is realized at $x$=0.7, and the stabilization
of AF1 state as function of hole-doping can be understood from the
doping-dependent modification of $J_1$ and $J_2$ coupling strength.

{\it Discussions}: We would like to address some important issues
here: 

(1) The proposed AF1 solution for large hole-doping side does
not break the 4-fold rotational symmetry. It is therefore expected
that the lattice distortion associated with the AF2 solution (the SDW
state in LaOFeAs, which break the 4-fold rotational symmetry) should
not occur for the AF1 phase region. 

(2) In the mean field theory, the quantum phase transition between AF1
and AF2 happens at $J_1=2J_2$. For a pure 2D system, however, the
spacial quantum fluctuation will completely destroy the long range
order in the AF2 phase, because of the absence of effective locking
terms between two different sub-lattices~\cite{JPHu}. On the other
hand, the long range order in AF1 state will survive under the quantum
fluctuation.  Therefore beyond mean field, the AF1 to AF2 transition
predicted here is actually replaced by AF1 to quantum disorder
transition. Turning on the inter-layer coupling will stabilize the
long range order for both AF1 and AF2 phases in finite
temperature. Due to the different behaviors of the quantum fluctuation
in AF1 and AF2 phases, the AF1 phase will be more classical with a
larger order parameter, while there will be still strong quantum
fluctuation in the AF2 phase, which greatly reduces the order
parameter and also may stabilize the superconducting phase around the
phase boundary. This scenario is also supported by the fact that the
observed ordered moment in BaFe$_2$As$_2$ (0.8$\mu_B$) is larger than
that in LaOFeAs (0.3$\mu_B$), and closer to the calculated one from
DFT, because of the enhanced 3-dimensionality in
BaFe$_2$As$_2$. Nevertheless, it should be kept in mind that the usage
of $J_1$-$J_2$ model is just approximate, and also it is not clear yet
how the on-site multi-orbital physics (inter-orbital fluctuation) may
modify the story.

(3) For both the electron and hole-doped sides, the observed
superconductivity phase region~\cite{BFS-SC} is located around the
QCP, although two QCPs are physically different.

(4) To justify our theory experimentally, it is important to have
systematical measurement of spin susceptibility as function of
hole-doping for high-quality single-crystals.

In summary, by first-principle calculation, we show that the
hole-doped side of FeAs-based compounds is very different with its
electron-doped side. For the high hole-doping side, we predict that
the checkboard-type AF1 state should be stabilized with partial
orbital-ordering but without lattice distortion. A unique AF2-to-AF1
QCP is realized with hole-doping, in sharp contrast to the
electron-doped side.

We acknowledge the valuable discussions with Z. Q. Wang, J. P. Hu, and
the supports from NSF of China and that from the 973 program of China
(No.2007CB925000).

\end{document}